# Analyzing the role of tensile and hybrid fractures on the maximum relief topography

**Julien Gargani** [1,2*]

[1] Université Paris-Saclay, Geops, CNRS, France
[2] Université Paris-Saclay, EST, France
* Correspondence: julien.gargani@université-paris-saclay.fr

**Abstract**

Slope stability description through mechanical laws has important implication for Earth morphology understanding and risk assessment. Previous researches have showed that shear, tensile, and hybrid fractures can be observed experimentally and in the field, but their descriptions by a single equation is still an open debate. Fracture envelope able to describe contemporaneously these three fracture modes differ significantly from the Mohr-Coulomb law. Despite the need to apply such a law at all scales, from the laboratory to the mountain range, the fracture criterion that characterizes all types of fractures is rarely used in geotechnical engineering and geological investigations. In order to analyze the stability thresholds of large-scale relief, the current work examines the effects of considering the Griffith criterion with a variable rock traction instead of the Mohr-Coulomb law using a modelling approach. The difference estimated on maximum relief using these two different rupture criterions could be of the same order than the one caused by geological phenomena, such as with or without seismic activity, or from the one caused by the destabilization processes (tilting vs. landslide). When compared to the modified Griffith criterion, the Mohr-Coulomb law tends to overestimate the maximum escarpment height. The results are examined in relation to the Carrara marbles, which serve as a case study for the theoretical framework.

---

## 1. Introduction

Landforms result not only from construction processes, but also from destructive processes [1,2]. Numerous destructive processes affect landforms. Destructive processes are of major geological interest for understanding long-term planetary surface evolution. These destructive processes are often triggered by gravitational forces. Gravitational processes are the main drivers of surface evolution, shaping landscapes through: (i) landslides [3-6], (ii) water erosion [7-13,101], (iii) glacial erosion [14], (iv) debris flow erosion [15-19,102], (v) rockfalls and rock tilting (i.e. toppling) [20-23], (vi) meteorite impacts [24]. Geochemical processes during rock alteration caused by fluid circulation may favor rocks weakening before rocks failure [25-27]. They constitute significant natural hazards with potential cascading effects, such as tsunamis, lake outburst floods, and vertical movements [28-32], and pose geotechnical challenges due to the difficulty to accurately predicting and mitigating them.

Destructive processes can result from a combination of several geological phenomena, such as slope destabilization caused by seismicity [33-34] or triggered by water infiltration and weathering [35-38]. Although these phenomena are often well-documented, distinguishing their respective contributions—particularly over long geological timescales or on distant planetary bodies—poses significant challenges.

Among the processes that cause the dismantling of landforms, the tilting of rock blocks are less studied than water erosion or landslides. However, slope destabilization through tilting represents a risk for populations and infrastructure [39-40]. Consequently, it is important to understand this mechanism and the uncertainties associated. The description of destabilization processes, particularly tilting, often relies on field approaches, remote sensing and modeling. Modelling approach consider physical laws, such as the equilibrium of forces and the failure criterion derived from rock mechanics experiments to describe the short and long term behavior under various geological and climatic conditions. Mechanical laws derived from experimental studies, such as the Mohr-Coulomb law [41] and the Griffith criterion [42], are widely used in the scientific literature as well as in engineering studies.

The Mohr-Coulomb law is the most often adopted failure criterion in geotechnical engineering [43-47]. It proposes a standardized approach, but leads to an oversimplification of the complex behavior of rock masses under tensile domain [48]. However, shear and tensile fractures can coexist in outcrops in the field [49]. The modelling of these two kind of fractures in mountain range, where they are observed, is necessary to describe the complexity of real world. Alternative criteria, such as the Griffith's law, or the use of a parabolic failure envelope [50], or the use of fracture envelopes with a quadratic equation [51,52], provide a more accurate description of fracture mechanics in the tensile and hybrid domains.

By comparing the maximum expected topographies according to the Mohr-Coulomb and Griffith models, we seek to determine which law best describes the thresholds of landscape stability at large scale. Small-scale mechanical properties should leave detectable imprints on the topography of high relief or entire mountain ranges. The relevance of experimental laws to describe what occurs at the scale of a mountain range is complex because of uncertainties related to the physical laws applied, but also of the superposition of different ways of destabilizing the landforms. The uncertainties surrounding instabilities are multifaceted and depend on: (i) triggering factors (e.g., seismicity [46,23], water infiltration [36,37], volcanic activity [53], (ii) the relative influence of different geological phenomena (landslide, water erosion, glacial erosion, freeze-thaw process) [34,4,53,54,55], and (iii) the choice of the mechanical law used for the analyses of the dynamic [56,30].

The observation of hybrid and tensile fractures experimentally [57] and on the field [58-61], and not only compressive fractures, is not well described by classical laws and could pose a problem for understanding the consequences of this type of fractures at the scale of mountain ranges. Can this property leave detectable imprints on the topography of hills and mountain ranges? This study addresses this question by comparing the maximum expected escarpment heights predicted using different failure criteria, in order to bridge the gap between laboratory-scale observations and large-scale geological phenomena. This study aims to explore the theoretical implications of adopting the Griffith criterion with variable traction for predicting instabilities, particularly on the tilting of massive rocks.

## 2. Methods and theoretical developments

### *3.1. Rock tilting modelling*

The forces acting on the tilting of the rock blocks are the weight $P$ and the resistance force of the rocks $R$. The weight is described by $P = \rho V g$, where $\rho$ the rock density, $V$ the rock volume and $g$ is the acceleration of the gravity. The rock resistance is described by $R = \tau \times S$, where $\tau$ is the maximum shear stress that the rock can resist before failure, and $S$ $= L \times W$ represents the surface area of the rock block expected to tip over in contact with the bedrock substrate. The volume of the block is $V = S \times H$, with $H$ the height of the rock block (Figure 1).

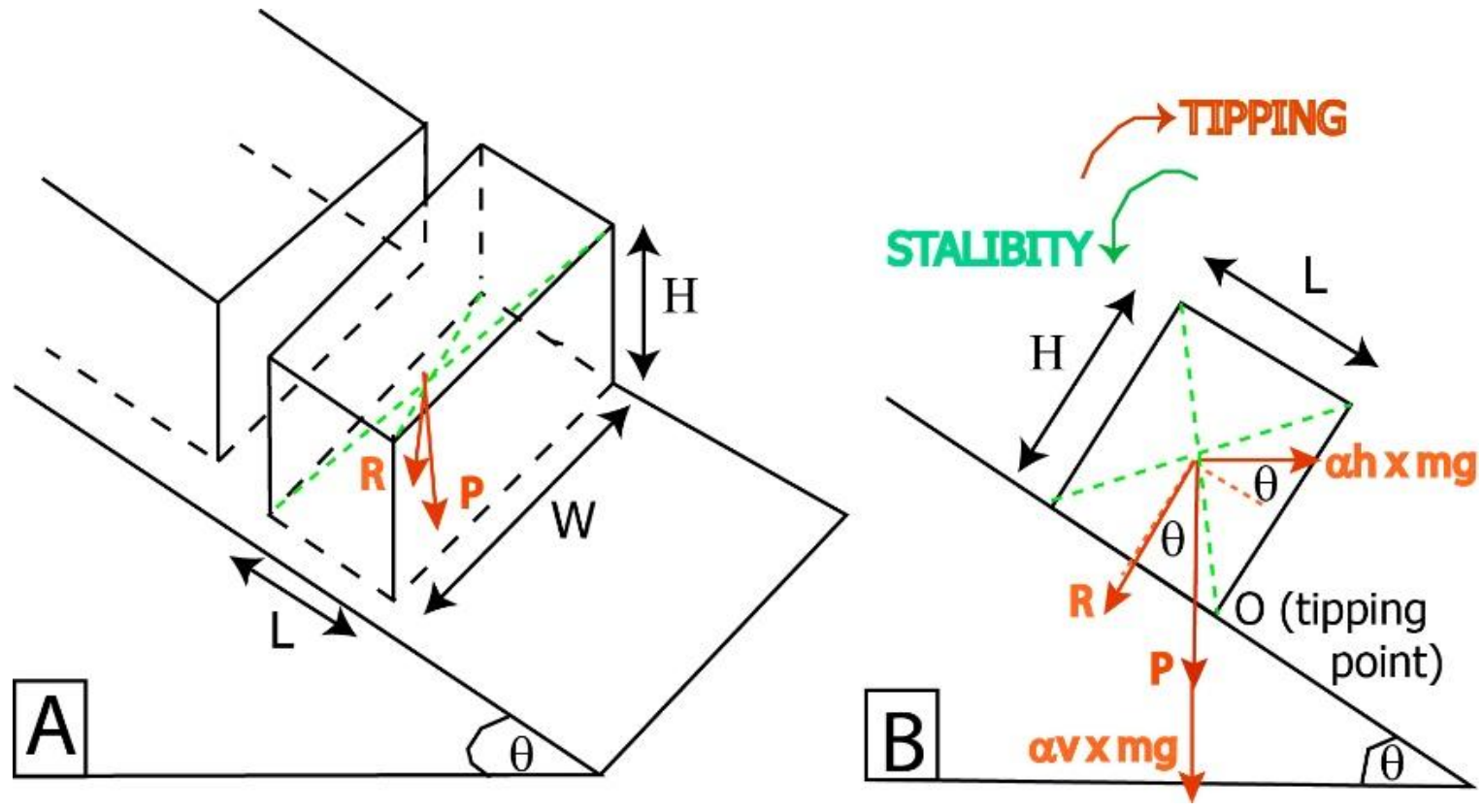

**Figure 1.** Schematic diagram of forces acting on rock block tilting. (A) 3D representation of the rock block, (B) 2D representation of the forces, including the forces triggered by the seismicity ($\alpha_h$mg and $\alpha_v$mg), acting on the rock block tilting.

The balance of moments of forces is given by the sum of all the moments of forces which oppose the tipping $\sum M_R$ on the one hand, and all the moments of forces which tend to tip over the block $\sum M_F$ on the other hand:

$$\sum M_R = \sum M_F$$

$$P_{\perp} \times L/2 + R \times L/2 = P_{//} \times H/2,$$

where $P_{\perp} = P \cos\theta$ is the normal component to the slope of the weight and $P_{//} = P \sin\theta$ is the tangential component to the slope of the weight, with $\theta$ the slope. Consequently,

$$\rho V g \cos\theta \times L/2 + \tau \times S \times L/2 = \rho V g \sin\theta \times H/2 \quad (1)$$

2.1.1. Mohr-Coulomb law

Considering that the maximum shear stress that the rock can resist is described by the Mohr-Coulomb law $\tau = \sigma_N \tan\phi + C$ where $\sigma_N$ is the normal stress, $\phi$ the angle of internal friction and $C$ the cohesion of the rock, the equation (1) can be written:

$$\rho \times SHg \cos\theta \times L/2 + (\sigma_N \tan\phi + C) \times S \times L/2 = \rho \times SHg \sin\theta \times H/2 \quad (2)$$

The normal stress $\sigma_N$ is obtained by the normal component of the weight:

$$\sigma_N = P_{\perp}/S = (\rho V g \cos\theta)/S = (\rho \times SHg \cos\theta)/S = \rho H g \cos\theta \quad (3)$$

Consequently, from (2) and (3), it can be written:

$$\rho \times SHg \cos\theta \times L/2 + (\rho \times Hg \cos\theta \tan\phi + C) \times S \times L/2 = \rho \times SHg \sin\theta \times H/2 \quad (4)$$

$$-H^2 \times [\rho g \sin\theta] + H \times (\rho \times Lg \cos\theta + \rho \times Lg \cos\theta \tan\phi) + LC = 0 \quad (5)$$

The solution of equation (5) is:

$$H = [\rho Lg (\cos\theta + \cos\theta \tan\phi) - \sqrt{[(\rho Lg)^2 (\cos\theta + \cos\theta \tan\phi)^2 + 4\rho \times LgC \sin\theta)]}] / [-2\rho g \sin\theta] \quad (6)$$

2.1.2. Griffith criterion

However, starting from equation (1), but assuming that the shear stress is described by the Griffith criterion:

$$\tau^2 = 4T\sigma_N - 4T^2 \quad (7)$$

where $T$ is the traction, it can be obtained that:

$$\rho \times SHg \cos\theta \times L/2 + \sqrt{(4T\sigma_N - 4T^2)} \times S \times L/2 = \rho \times SHg \sin\theta \times H/2 \quad (8)$$

$$\rho Hg \cos\theta \times L/2 + \sqrt{(4T\rho Hg \cos\theta - 4T^2)} \times L/2 = \rho Hg \sin\theta \times H/2 \quad (9)$$

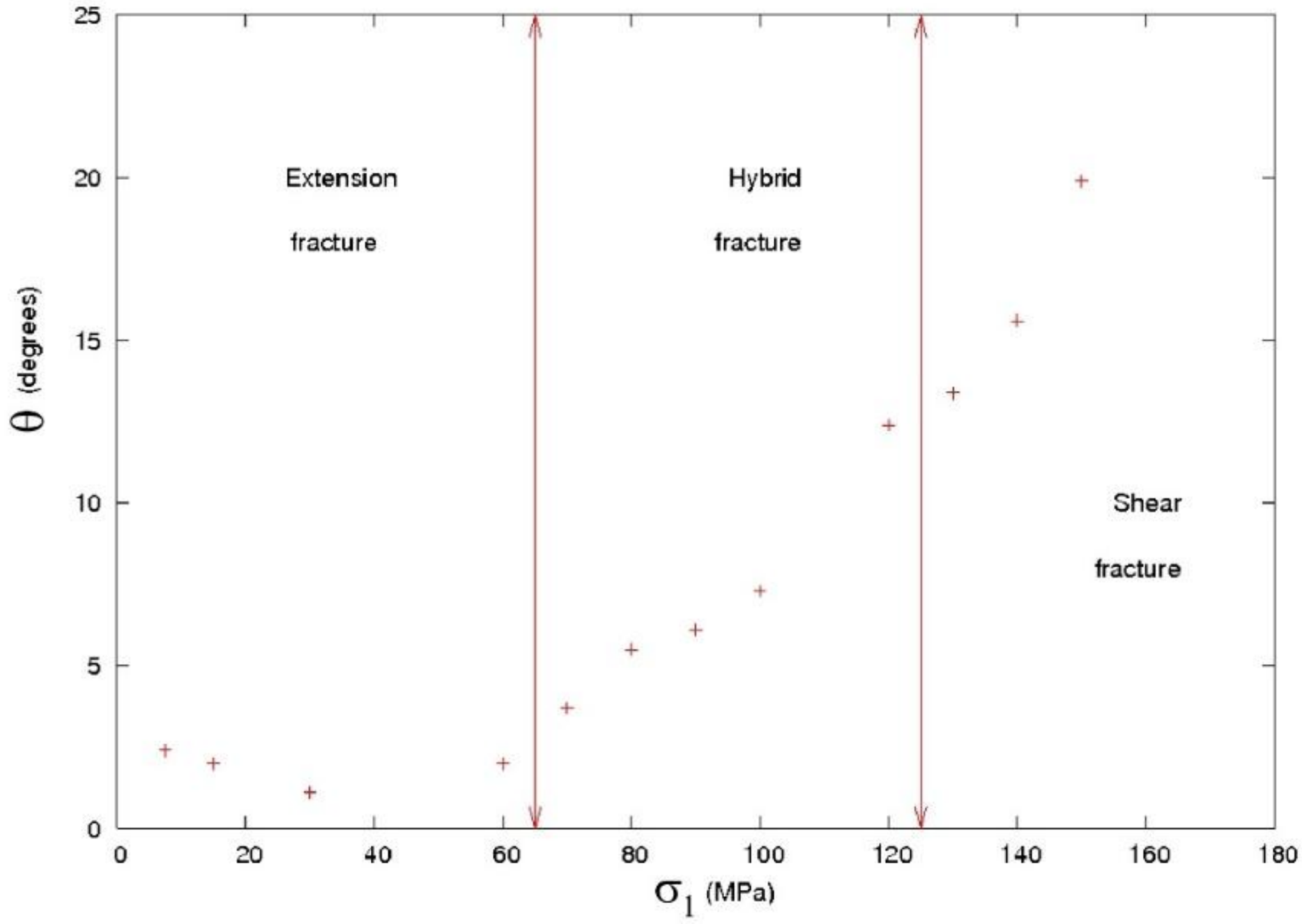

**Figure 2.** Fracture angle θ as a function of $\sigma_1$ for Carrara marbles. The shear, hybrid and tensile fracture domains are indicated. Data from Ramsey and Chester (2004) [57].

From equation (9), it can be obtained that:

$$H^4 [(\rho g \sin\theta/(2L))^2] - H^3 [(\rho g/2)^2 \times 2(\sin\theta \cos\theta)/L] + H^2 [(\rho g \cos\theta)/(2L))^2] - H [T \rho g \cos\theta] - T^2 = 0 \quad (10)$$

This quadratic equation is solved numerically using the Ferrari algorithm implemented in Fortran. There are four real solutions and the real part of each solution is presented in the result section.

*2.2. Resistance to rock traction*

2.2.1. Derivation of empirical results

Experimental results show that shear, tensile and hybrid fractures take place [57,51] (Figure 2). These results are obtained using experimental devices conducting to obtain a set of data ($\sigma_1$,$\sigma_3$,θ) for each fracture.

These sets of data could be converted in shear stress $\tau$ and normal stress $\sigma_n$ using a geometrical analysis [62]:

$$\sigma_n = \sigma_1 \times (1 - \cos 2\theta)/2 + \sigma_3 \times (1 + \cos 2\theta)/2 \quad (11)$$

$$\tau = \sigma_1 \times (\sin\theta/\cos\theta) + \sigma_3 \times (\cos\theta/\sin\theta) \quad (12)$$

For example, using the experimental values of $\sigma_1$, $\sigma_3$, $\theta$ obtained by Ramsey and Chester (2004) [57], the values of the shear stress $\tau$ and of the normal stress $\sigma_n$ can be represented on the same graph as in the Mohr classical representation. These data show a discrepancy from the Mohr-Coulomb law, but also from the Griffith criterion (Figure 3).

2.1.1. Variation of the resistance of rock traction

It can be associated at each experimental values of $(\sigma_1, \sigma_3, \theta)$ a value of $(\tau, \sigma_n)$. Consequently, using equation (7), it can be obtained a value of the rock traction $T$. The rock traction $T$ can be represented as a function of $\sigma_n / \tau$ or as a function of $\sigma_3 / (\sigma_1 \times \theta)$ (Figure 4). $T$ is not always constant under these representations. More precisely, under extensive conditions when $\tau \approx 0$ MPa or when $\sigma_n / \tau < -4$ MPa, the rock traction is constant with $T \approx 7.8$ MPa. However, when $\sigma_n / \tau > -3$ MPa the rock traction T is not constant. $T$ can be described by a function such as (Figure 4):

$$T = T_0 + e^{a(\sigma_n/\tau)} + b\, e^{-((\sigma_n/\tau)+c)^2} \tag{13}$$

*or*

$$T = T_0 + e^{d(\sigma_1/(\theta\sigma_3))} + f\, e^{-g((\sigma_1/(\theta\sigma_3))+h)^2} \tag{14}$$

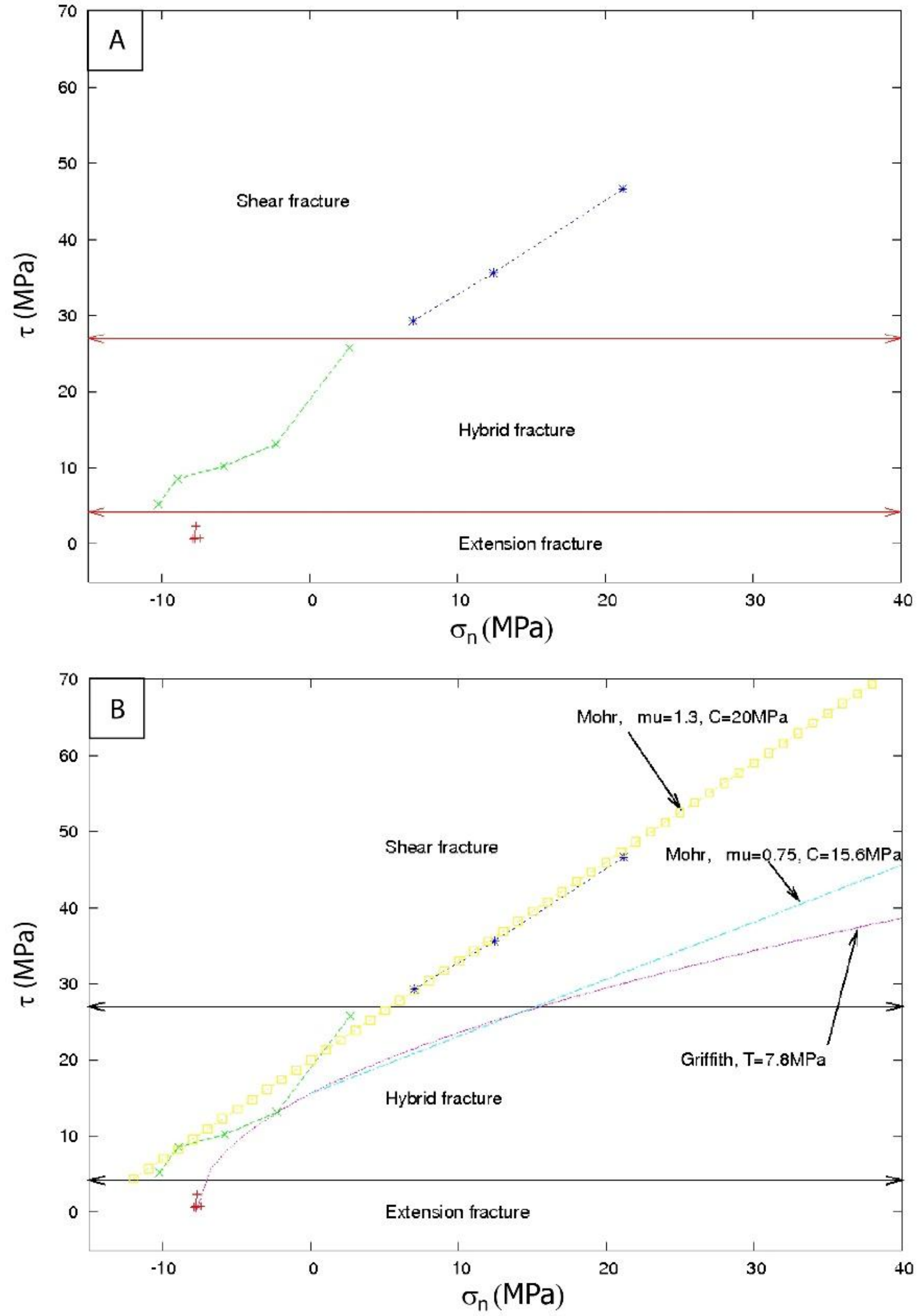

**Figure 3.** Shear stress τ and normal stress $\sigma_n$. (A) Experimental results are obtained from the data from Ramsey and Chester (2004) [57], (B) The Mohr-Coulomb law is able to fit the shear and hybrid fractures, assuming that mu = tanϕ = 1.3 and C = 20 MPa (yellow line), but not the extensional fractures in the tensile domain. The Griffith criterion is not able to fit all the data contemporaneously. T=7.8 MPa (Pink line) [57].

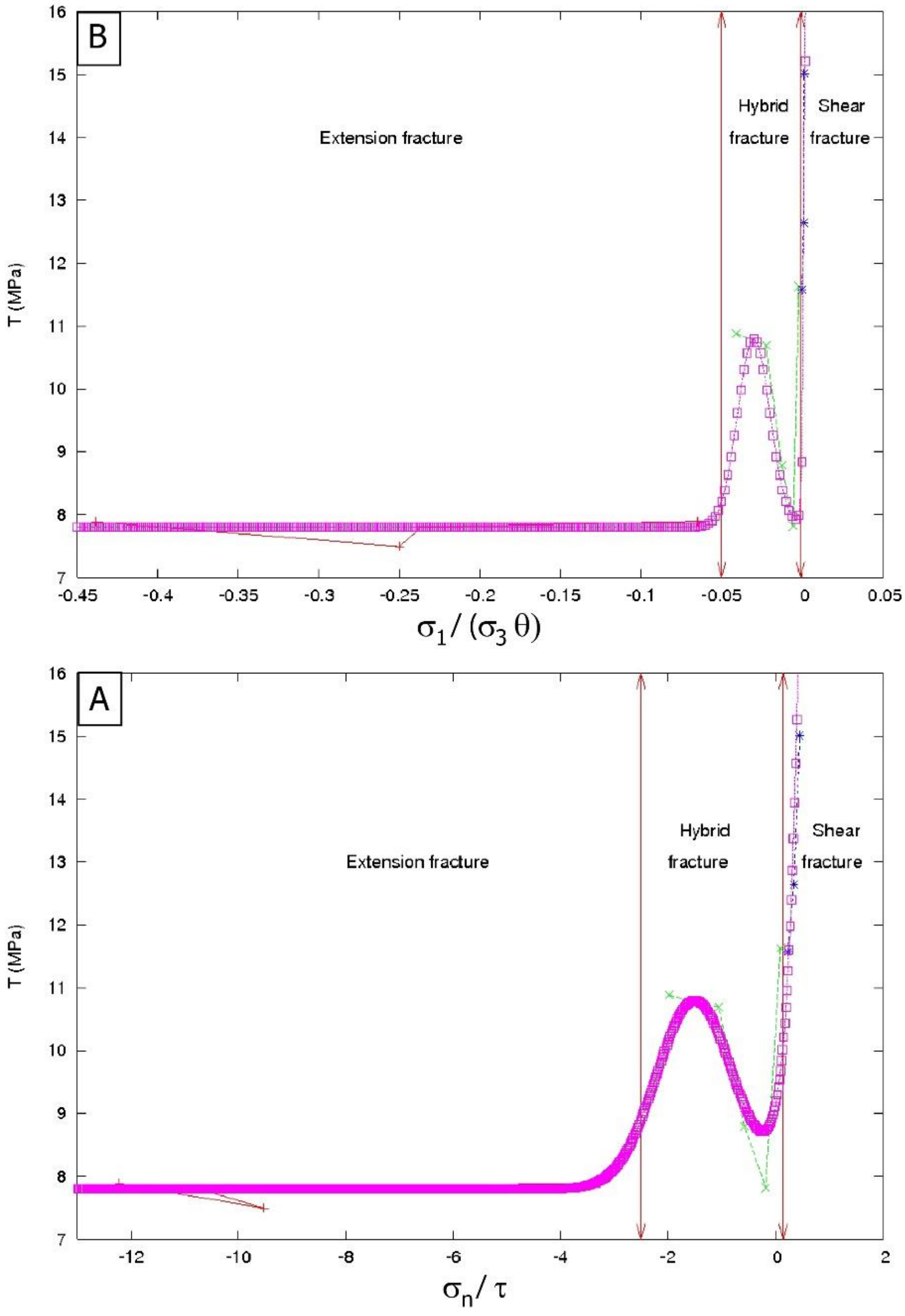

**Figure 4.** Theoretical and experimental rock traction T represented (A) as a function of $\sigma_n / \tau$ as in equation 13 or, (B) as a function of $\sigma_3 / (\sigma_1 \times \theta)$ as in equation 14. $T_0$=7.8 MPa, a=5, b=3 MPa, c=1.5, d=1000, f=3 MPa, g=5000, h=0.003. Experimental data are from Ramsey and Chester (2004) [57].

2.2.3. Fitting the contemporaneously shear, hybrid and extensive fractures

The Mohr-Coulomb equation $\tau = \sigma_N \tan \phi + C$ can represent adequately the shear fracture and hybrid fracture, considering than $\tan \phi = 1.3$ (Figure 3). However, the Mohr-Coulomb is not appropriate to fit contemporaneously the shear fractures, the hybrid fractures and the tensile fractures. The Griffith criterion also fail to interpret contemporaneously all the kinds of fractures considering a constant traction *T*. The Griffith criterion can be used to describe the Carrara marbles considering a variable traction *T* (Figure 4B, equation 13).

Another way to describe the fracture envelope is to fit the data using exponential and quadratic functions. The shear, extensive and hybrid fractures can be better fitted using equations (15) and (16) (Figure 5 and figure 6):

$$\tau = 1 + z \times e^{\sigma_n/(m\tau) + n/m)} \quad (15)$$

$$\sigma_n = k \times (\tau + p)^2 - q \times (\tau + p) + r \times e^{u\tau^2} - w^2 \quad (16)$$

where *m*, *n*, *k*, *p*, *q*, *r*, *u* and *w* are empirical constant.

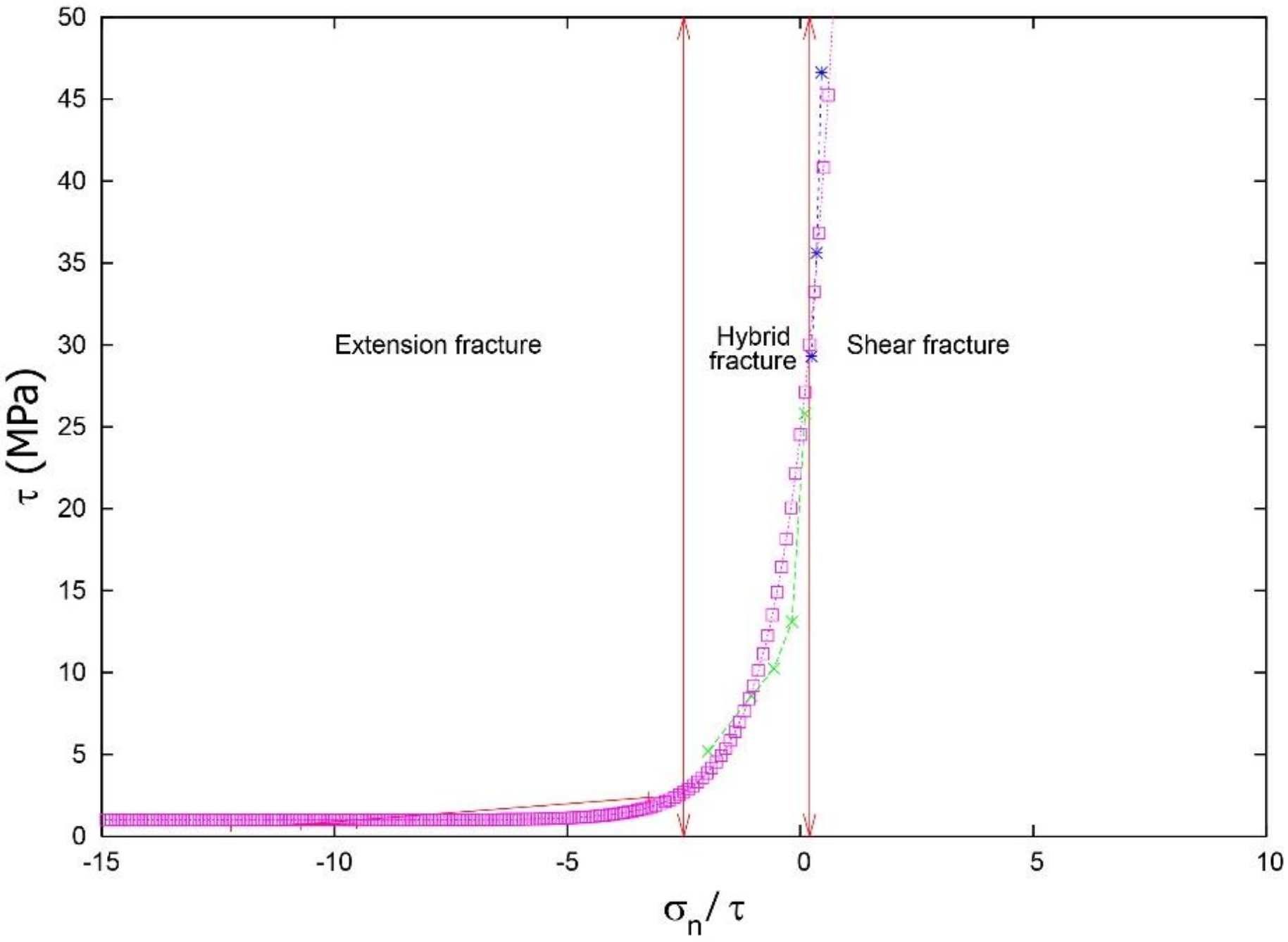

**Figure 5.** Theoretical and experimental fracture threshold of the shear stress τ as a function of $\sigma_n$ / τ for Carrara marbles. z=1 MPa, m=0.95, n=3. Pink curve represents $\tau = 1 + e^{\sigma_n/(0.95\tau) + 3/0.95)}$. Experimental data are from Ramsey and Chester (2004) [57].

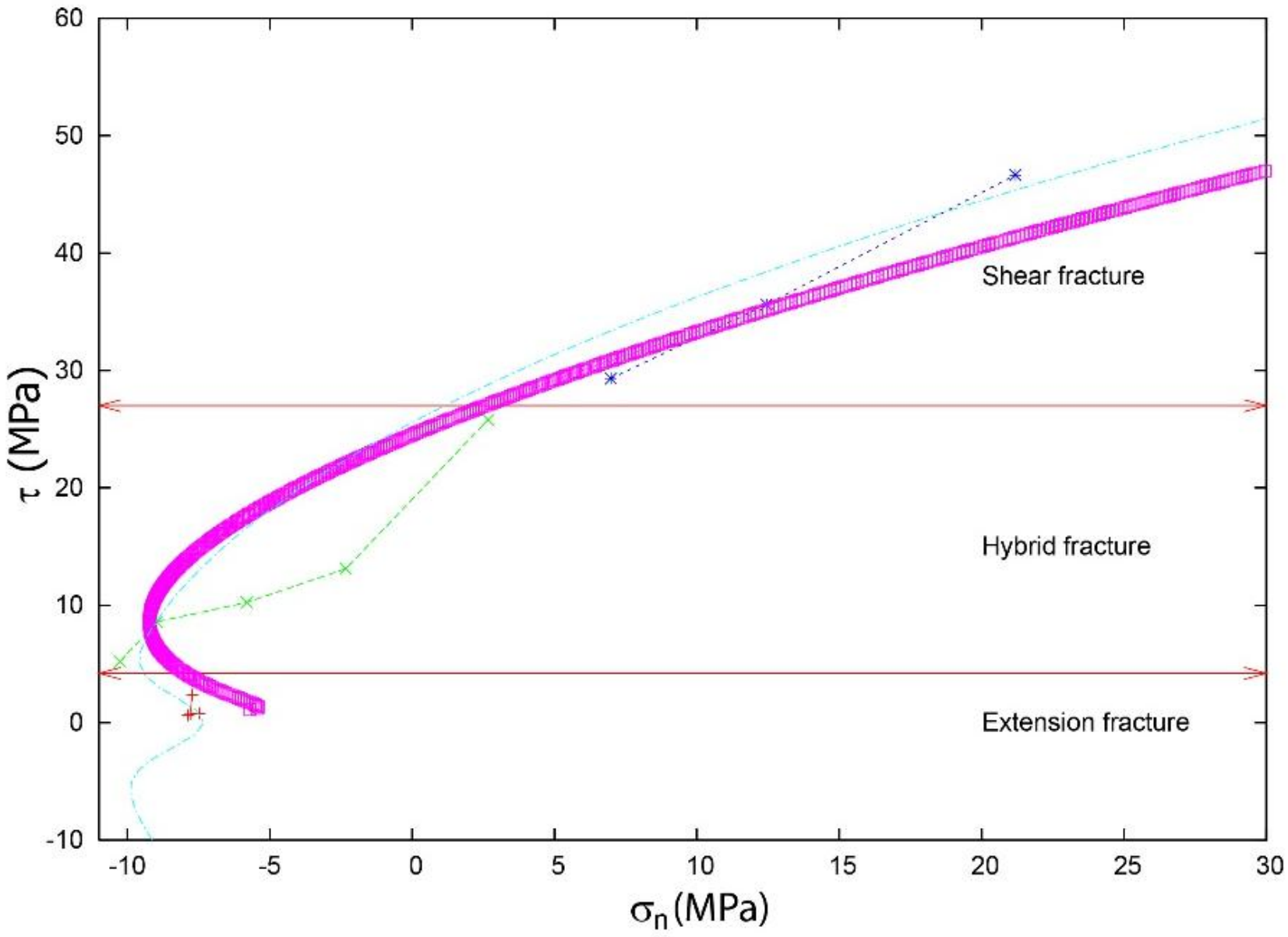

**Figure 6.** Theoretical and experimental fracture threshold of the shear stress τ as a function of $\sigma_n$ for Carrara marbles. k=0.0147 MPa$^{-1}$, p=20 MPa, q=0.56, r=3 MPa, u=0.1 MPa$^{-2}$, w=5 MPa. $\sigma_n = 0.0147 \times (\tau + 20)^2 - 0.56 \times (\tau + 20) + 3.e^{0.1\tau^2} - 5$. Experimental data are from Ramsey and Chester (2004) [57].

In order to implement the new equations of shear stress and normal stress to estimate the relief stability in the case of Griffith criterion with a variable traction *T*, the calculation of the maximum relief *H* with the slope $\theta$ is estimated using the following workflow: (i) Starting from an initial constant value of the traction $T_0 = 7.8$ MPa, for all the slope $\theta$ values, (ii) calculate the initial maximum relief $H_0$ of the quadratic equation (10) using the Ferrari method for $T=T_0$. By this way, (iii) the initial normal stress $\sigma_{n0} = \rho\, g\, H_0 \cos\theta$ can be calculated as well as initial shear stress $\tau_0 = \sqrt{(4T_0\, \rho H_0 g \cos\theta - 4T_0^2)}$, corresponding to equation (7). Then, (iv) $\sigma_{n1}$ and $\tau_1$ are calculated using equations (16) and (15), respectively. (v) The ratio $\sigma_{n1} / \tau_1$ is estimated. (vi) then the traction $T_1$ is calculated using equation (13). Finally, (vii) the maximum relief $H_1$ obtained from the quadratic equation (10) is calculated using the first solution from the Ferrari method for $T=T_1$.

## 3. Results

The shape of the failure envelope is a shifted parabola opening at the right (Figure 6) as observed in previous studies [51,52]. The new equation (equation 16) used to model the behavior of rocks, including the tensile and hybrid domain, influence significantly on the estimate of the maximum height of escarpment.

### *3.1. Rock tilting stability threshold using the Mohr-Coulomb law*

Using equation (6), the maximum relief *H* is obtained for a slope $\theta$. In this case, the relief is considered to be destabilized only by the tilting of the massive block considered and by no other process. Each line of figure 7 represents a threshold of stability for a given length of the rock block. Above the lines, the relief is not stable, and is expected to collapse, for a given length *L* of the bloc. Below the line, the relief is expected to be stable. When the slope $\theta$ increases, stability decreases and a lower relief height is expected. Using the Mohr-Coulomb law to assess the stability of a landform caused by tilting, the results show that the longer the rock block *L* is, the more stable the rock block is. It can be observed that when *L* is sufficiently high (*L*>50-100m), the ratio *H*/*L* is almost identical when *L* increases. From a theoretical point of view, the shape ratio *H*/*L* became rapidly an invariant and is not influenced by *L*, when *L* is high (L>50-100m). In the following numerical experiments, a reference curve is considered for comparison for L value causing a ratio *H*/*L* invariant (L=1000m, red line).

### *3.2. Influence of climate and seismicity on rock block tilting threshold*

Quantification of the influence of water on the relief stability, can be done considering that the Mohr-Coulomb law is written $\tau = (\sigma_N - U) \times \tan\phi + C$, where *U* is the pore pressure. The presence of pore pressure modifies equation (5) by adding a new term. To determine the maximum height of escarpment *H*, it is now necessary to solve the equation:

$$-H^2 \times [\rho g \sin\theta] + H \times (\rho L g \cos\theta + \rho L g \cos\theta \tan\phi) + L \times (C - U \tan\phi) = 0 \qquad (17)$$

The increase in pore pressure causes a lowering of the maximum height of the escarpment (Figure 8, green curve to be compared to the red curve). This means that humid climatic conditions or fluid circulation could destabilize a relief that is close to equilibrium. This result is in accordance with previous studies [63,64].

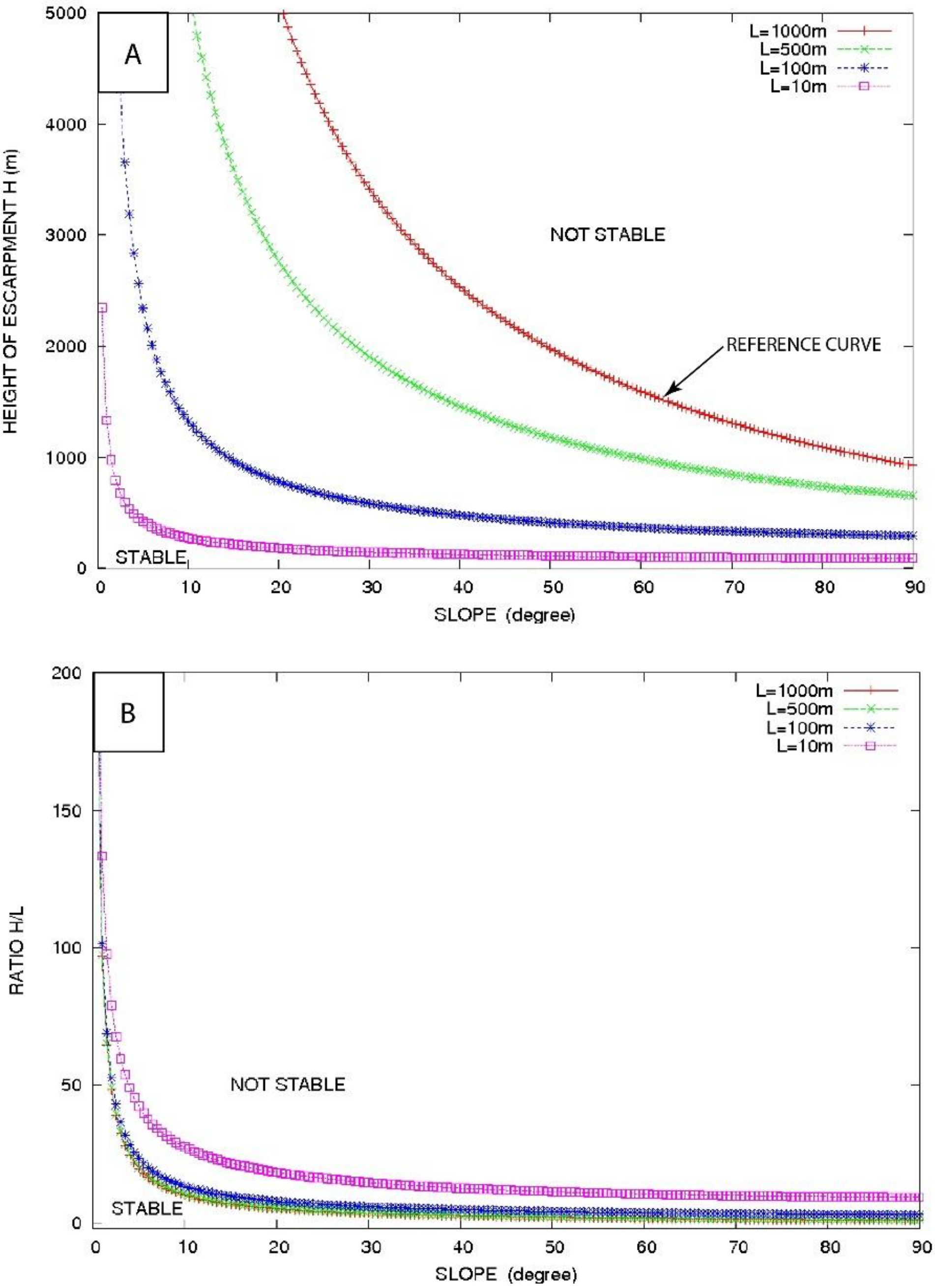

**Figure 7.** Tilting threshold. Rock tilting threshold for different geometries using the Mohr-Coulomb law. The cohesion is C=25.2MPa, the angle of internal friction is $\phi$=34.3, g=9.8ms$^{-2}$, $\rho$=3000kg/m$^3$, the rock length L is variable. (A) The threshold of the maximum height of escarpment H is indicated in meters; (B) the ratio between the maximum height of escarpment H and the length L is represented. The curves for L = 100 m, 500 m and 1000 m are superimposed, suggesting that almost similar threshold ratio H/L are estimated for a given slope θ, when L became sufficiently high (>50-100m).

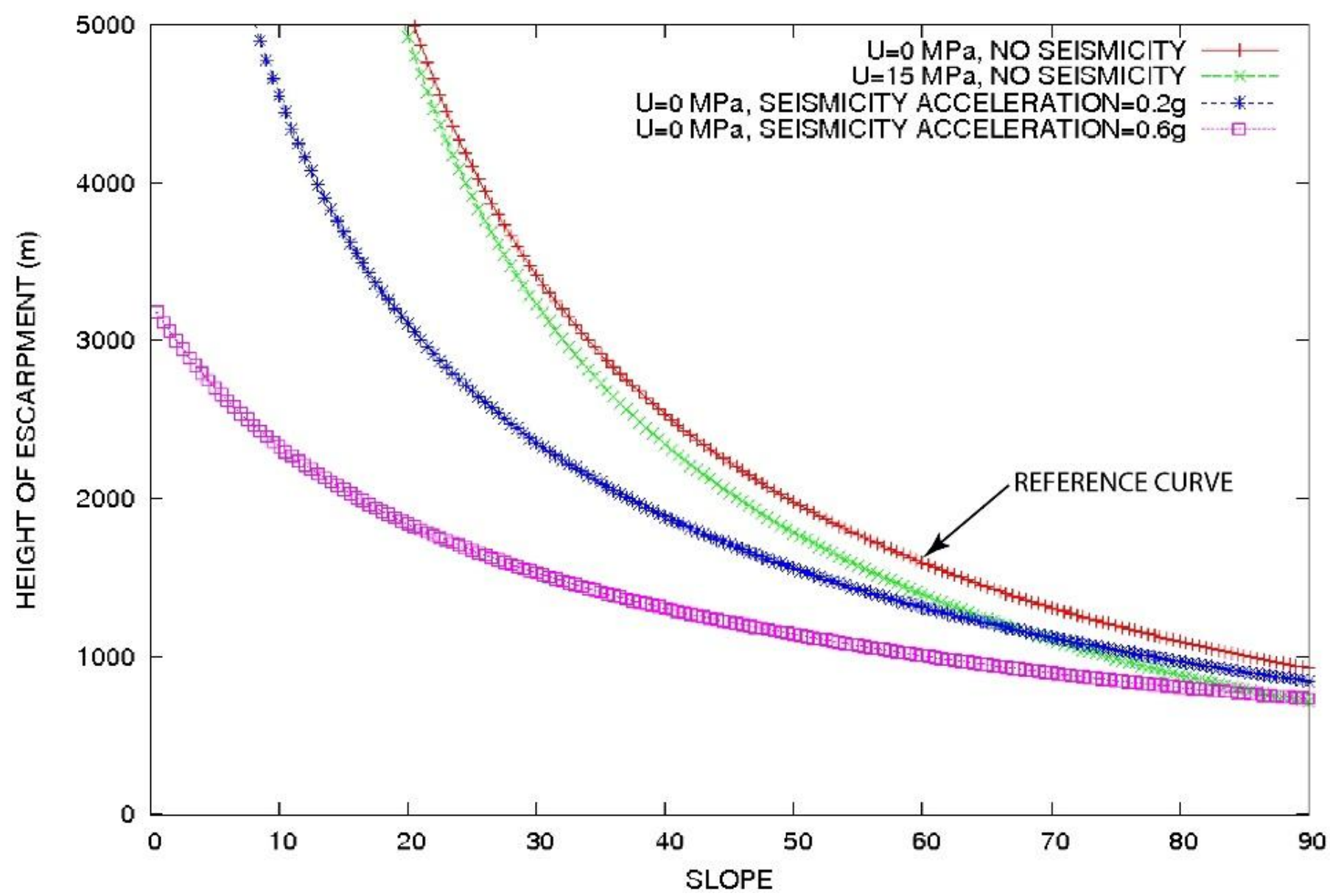

**Figure 8.** Tilting threshold. Influence of pore pressure and seismicity on relief stability to tilting. Coefficient of ground acceleration αg is $\alpha_H = \alpha_V = 0.2$ (blue line), $\alpha_H = \alpha_V = 0.6$ (pink line), the pore pressure U=15 MPa (green line), C=25.2MPa, ϕ=34.3, g=9.8ms$^{-2}$, ρ=3000kg/m$^3$, L=1000m. The reference curve visible in figure 7 is indicated.

Another well-known cause of destabilization of relief is seismicity. Seismic forces that favor vertical and horizontal movements could be written $F_V=\alpha_V mg$ and $F_H=\alpha_H mg$, respectively, where $\alpha_V$ and $\alpha_H$ are coefficient expressing a modification of the acceleration, leading to consider the peak ground acceleration $\alpha g$ (PGA) as a useful parameter [65]. The higher the seismicity, the bigger the coefficients $\alpha_V$ and $\alpha_H$. These forces modify the force balance conducting to a new expression to the equation to solve to obtain $H$:

$$-H^2 \times [\rho g\,(\sin\theta + \alpha_V \cos\theta + \alpha_H \sin\theta)] + H \times (\rho L g \cos\theta + \rho L g \cos\theta \tan\phi) + L\,C = 0 \quad (18)$$

When $\alpha_V$ and $\alpha_H$ increase, the maximum height of the escarpment decreases (Figure 8, blue curve, to be compared to the red curve). This result is in accordance with the finding by previous studies [34].

### *3.3. Rock sliding vs rock tilting threshold for massive relief using the Mohr-Coulomb law*

When the rock block is relatively small (i.e. $L \approx 10$m), the rock tilting process is easier to be observed than the rock sliding at higher slopes. However, for large reliefs lentgth (i.e. $L \approx 1000$m, red reference curve) tilting is more difficult to achieve than sliding for higher slopes (>60°) (Figure 9). At the opposite, the tipping threshold is lower than the sliding threshold for the gentler slopes (<34°).

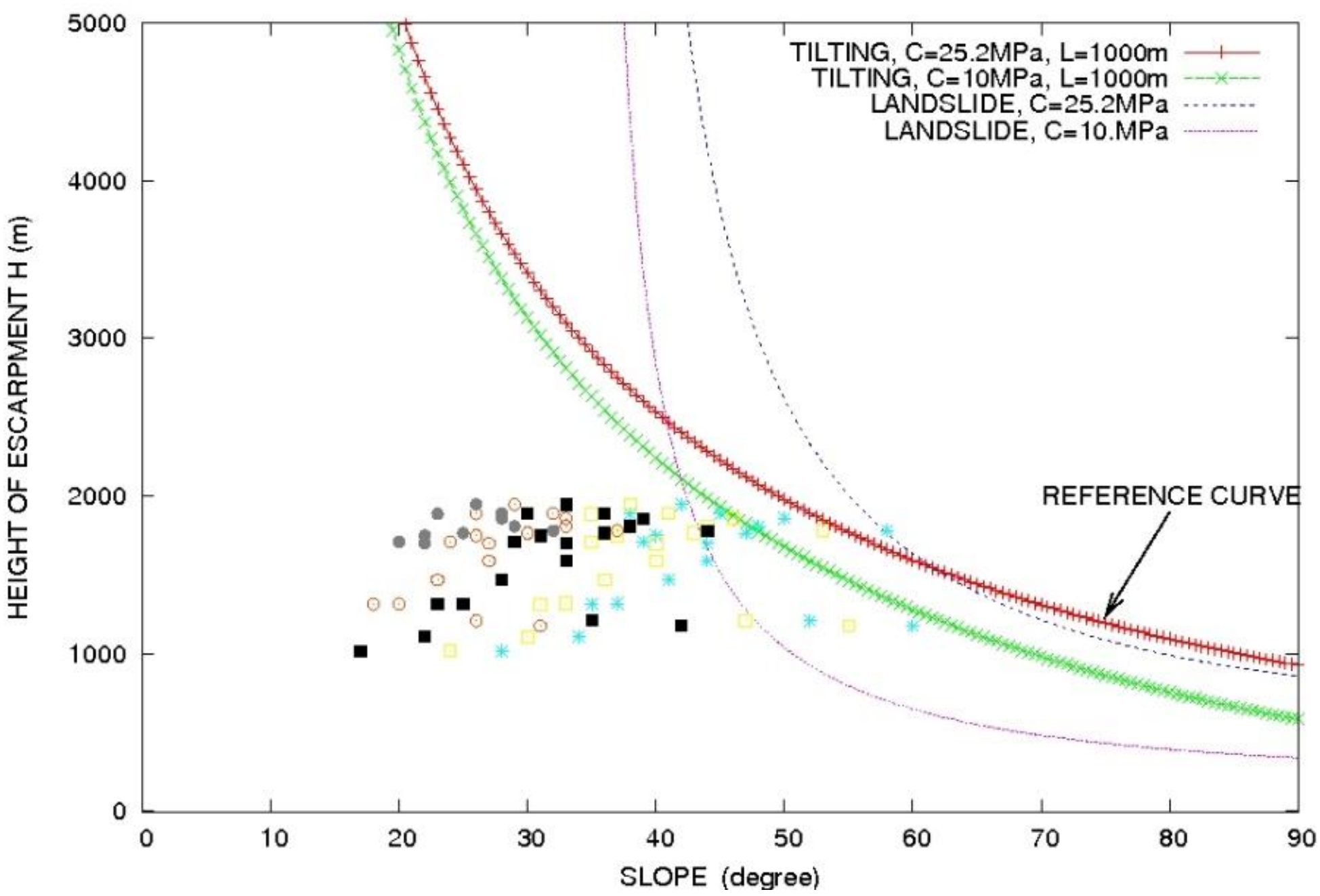

**Figure 9.** Tilting and sliding thresholds. Comparison of the slope stability for landslide versus rock fall using Mohr-Coulomb law. ϕ=34.3°, C=25.2MPa, g=9.8ms$^{-2}$, ρ=3000kg/m$^3$, L=1000m for tilting. Observed relief features in the Apennines mountain range in the area of the Carrara marbles lithology are indicated. The reference curve is the same as in figure 7 and 8.

*3.4. Rock block stability threshold using the Griffith law*

Using the Griffith's law to assess the stability of a landform subjected to tilting, the result shows that there is more than one solution (Figure 10). There are three positive solutions and one negative. The negative solution is not shown. Two of the solutions are identical. Consequently, there is only two different solutions. One of these solution is equal to zero when $\theta$=90° and is independent from $T$. It can be observed in figure 9 that the higher is the slope $\theta$, the less stable the relief $H$ is.

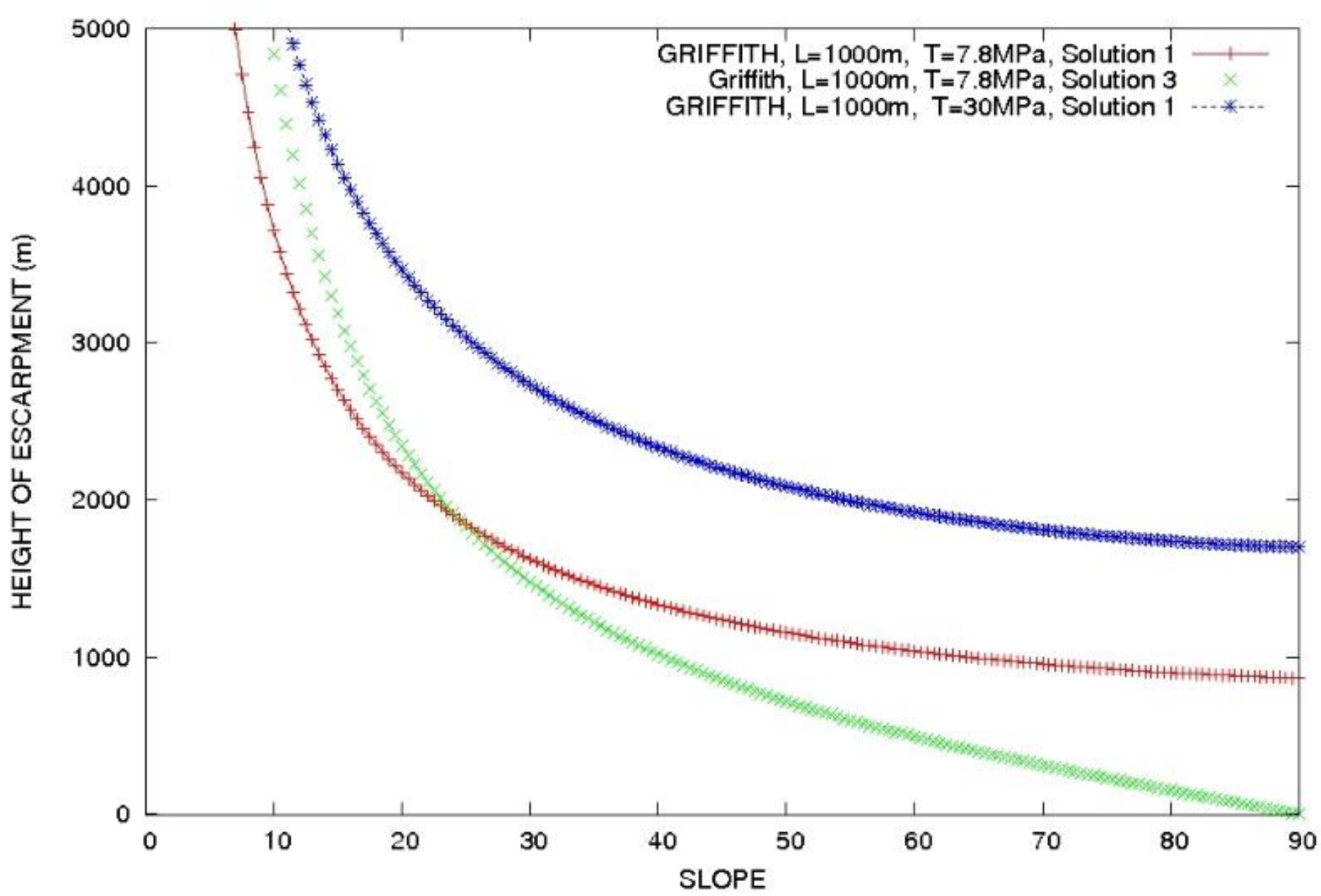

**Figure 10.** Tilting threshold. Effect of the Griffith criterion with a variable traction T on the tilting stability. There are two different solutions for the maximum height of escarpment H for a given variable traction T. One of these solution is independent from T (green curve, called solution 3). g=9.8ms$^2$, ϕ=34.3, L=1000m, ρ=3000kg/m$^3$, a=5, b=3 MPa, c=1.5, $T_0$=7.8MPa and $T_0$=30MPa.

### *3.5. Comparing rock block tilting stability threshold: Mohr-Coulomb vs Giffith*

The thresholds calculated using the Mohr-Coulomb and Griffith laws have all a concave shape (Figure 11). There is a solution obtained using the Griffith's law (blue line, *T*=30MPa) which is closer to the solution obtained with Mohr-Coulomb's law, for the same block length (here, *L*=1000m). However, the value *T*=30MPa is very different from experimental values (factor 4). The difference between the Mohr-Coulomb and the Griffith method is reduced when considering a factor of 2 for the length (or a factor of 2 for the distance from the tilting point).

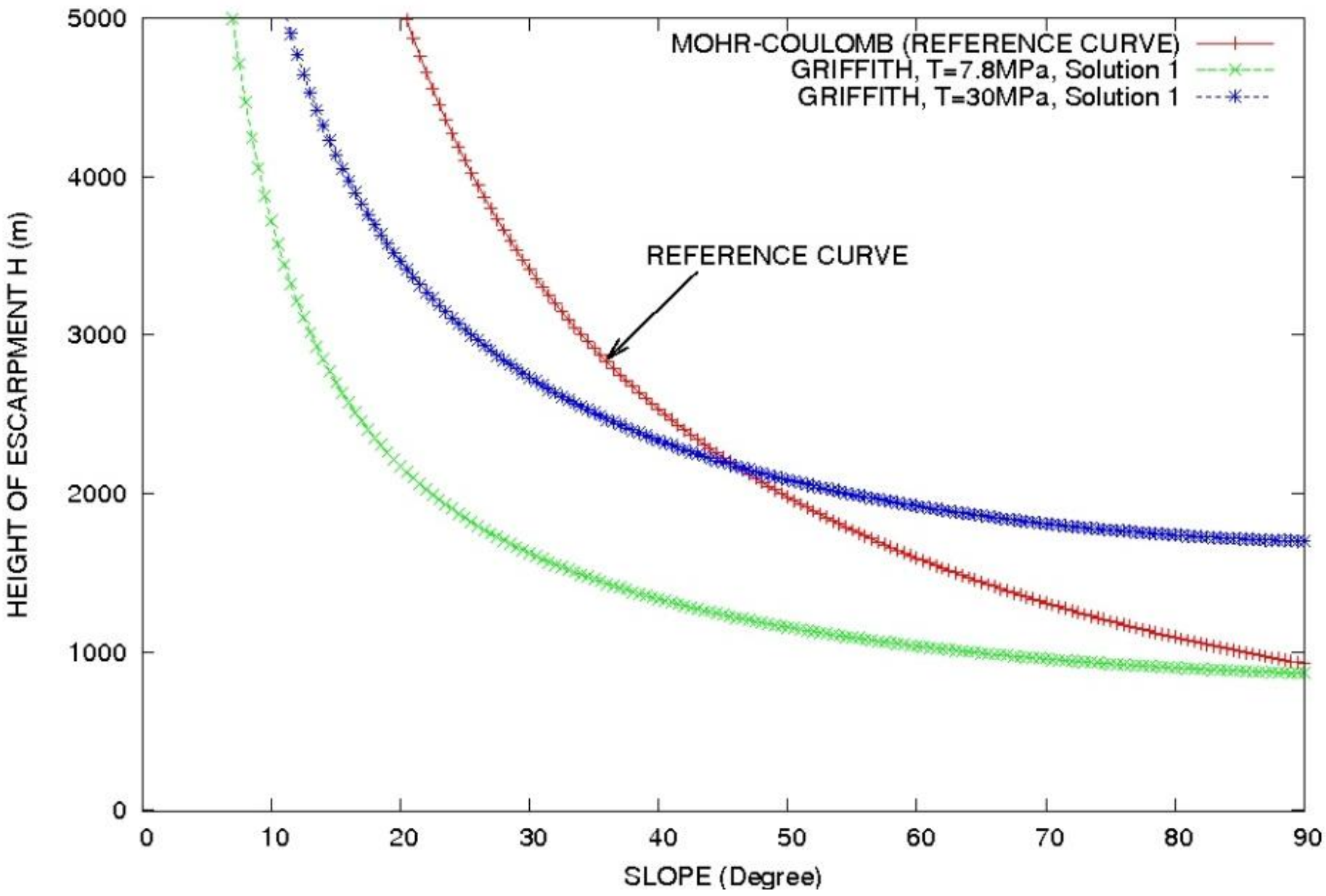

**Figure 11.** Tilting threshold. Comparison of the effect of the Mohr-Coulomb and Griffith laws, with a constant traction T, on the stability of relief H to tilting as a function of the slope θ. The reference curve visible in figure 7, 8 and 9 is indicated. g=9.8ms$^{-2}$, C=25.2MPa, ϕ=34.3, ρ=3000kg/m$^3$, $T_0$=7.8MPa and $T_0$=30MPa, a=5, b=3 MPa, c=1.5. For the Griffith law, with a constant traction T, only the positive solution that depends from the traction value T is represented.

### *3.6. Giffith Influence of the effective resistance to rock traction on relief*

The difference observed from the Mohr-Coulomb reference curve is significant for the Griffith curves with a constant traction *T*. These differences are comparable to the effect caused by seismicity (Figure 8) or when other destabilization processes are dominant, as landslide (Figure 9).

The semi-theoretical approach developed in this study suggests that the effective traction *T* is not constant (equation 13). From an analysis based on the balance of forces, and considering the Griffith failure criterion with a variable traction *T* as the law describing the rock behavior, the maximum height of escarpment is calculated. Thus, the maximum height of escarpment calculated according to the Griffith criterion ($H_G$ with traction *T* variable, pink curve) has a shape closer to that estimated by the Mohr-Coulomb law ($H_{M-C}$, brown thin curve) than previously ($H_G$ with traction *T* constant, green curve) (Figure 12). The maximum height of escarpment calculated using the Griffith criterion is lower than the one calculated using Mohr-Coulomb, i.e. $H_G < H_{M-C}$. However, considering in equation (13) that $T_0$ is equal to 11 MPa instead of 7.8 MPa is sufficient to obtain very similar results between $H_G$ (light blue curve) and $H_{M\text{-}C}$ (red thin curve). The possibility to fit the reference curve using the Griffith criterion with a variable traction $T$ suggests that $C$ and $\phi$ effective values for the Mohr-Coulomb parameters could be different from the measured ones.

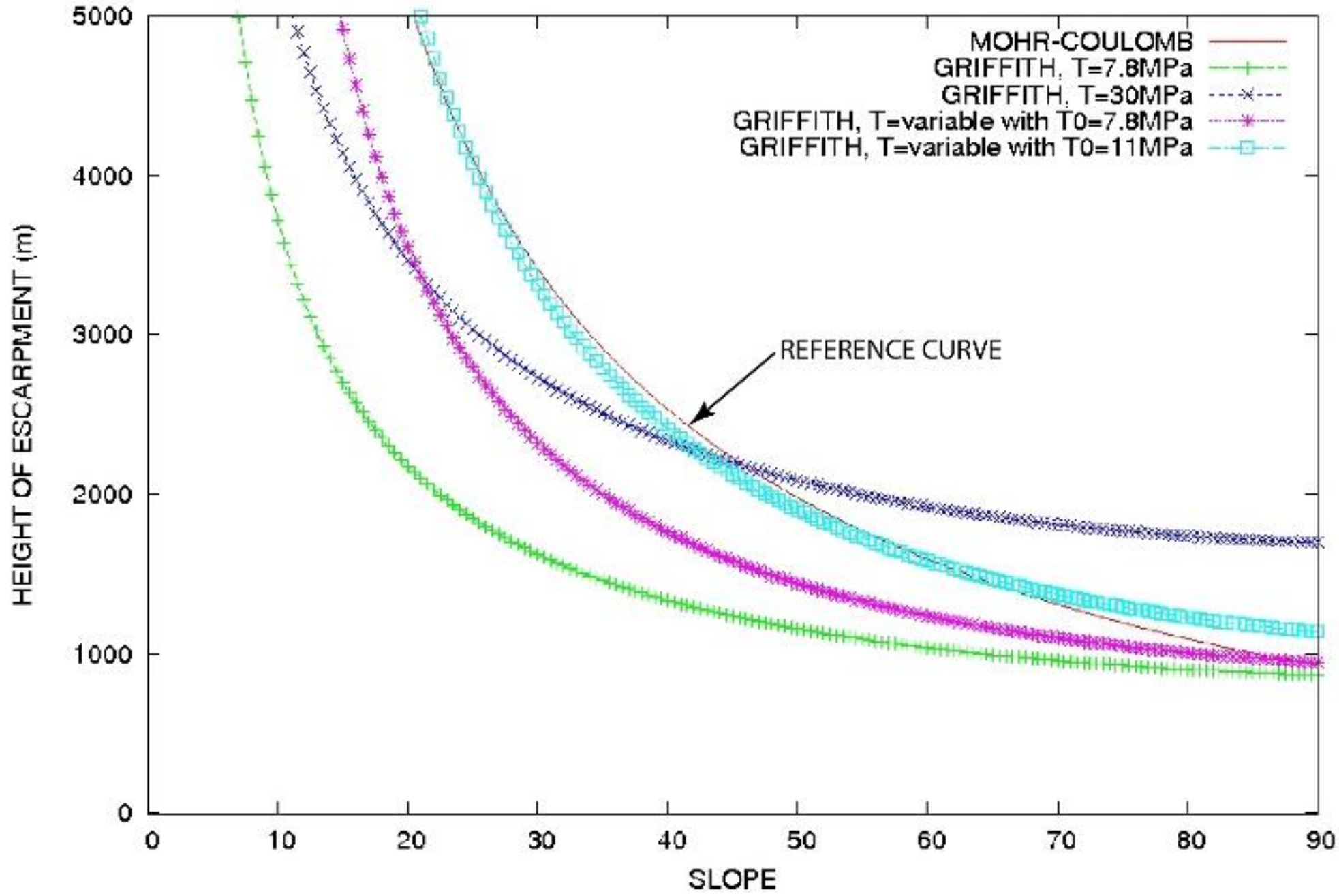

**Figure 12.** Effect of the variation of $T$ on the bloc tilting stability. Considering a variable T for the Griffith model conduce to a higher similarity with the Mohr-Coulomb method. $C$=25.2 MPa, $\phi$=34.3°, $L$=1000m, $g$=9.8ms$^{-2}$, $\rho$=3000 kg/m$^3$, $a$=5, $b$=3 MPa, $c$=1.5. The thin line is the reference curve visible in figure 7, 8, 9, 11.

## 4. Discussion

The influence of: (i) geological phenomena, such as seismic activity or climatic variations [66,67], (ii) the combination of various slope destabilization processes, such as landslides, tilting, water erosion, or glacial erosion [68,69], is not negligible on morphology. Here, it has been shown that their influence is of the same order of magnitude as the failure criteria used to estimate the maximum height of an escarpment.

### *4.1. Discriminating between the main destabilization processes at the mountain range scale*

From a theoretical point of view, it is possible to distinguish between sliding and tilting as the main process of slope destabilization, by analyzing the maximum height of the escarpments in a given place (Figure 13). This in in agreement with the results suggested in previous studies [70,63]. Although such a distinction is theoretically possible, its implementation using only altitudinal data is more complex due to the multiple processes that destabilize the relief. Indeed, other processes, such as glacial erosion [14], runoff erosion [71], landslides and debris flows [72], simultaneously influence the maximum elevations reached by the relief by continuously destabilizing the slopes in the Carrara area during the last million years.

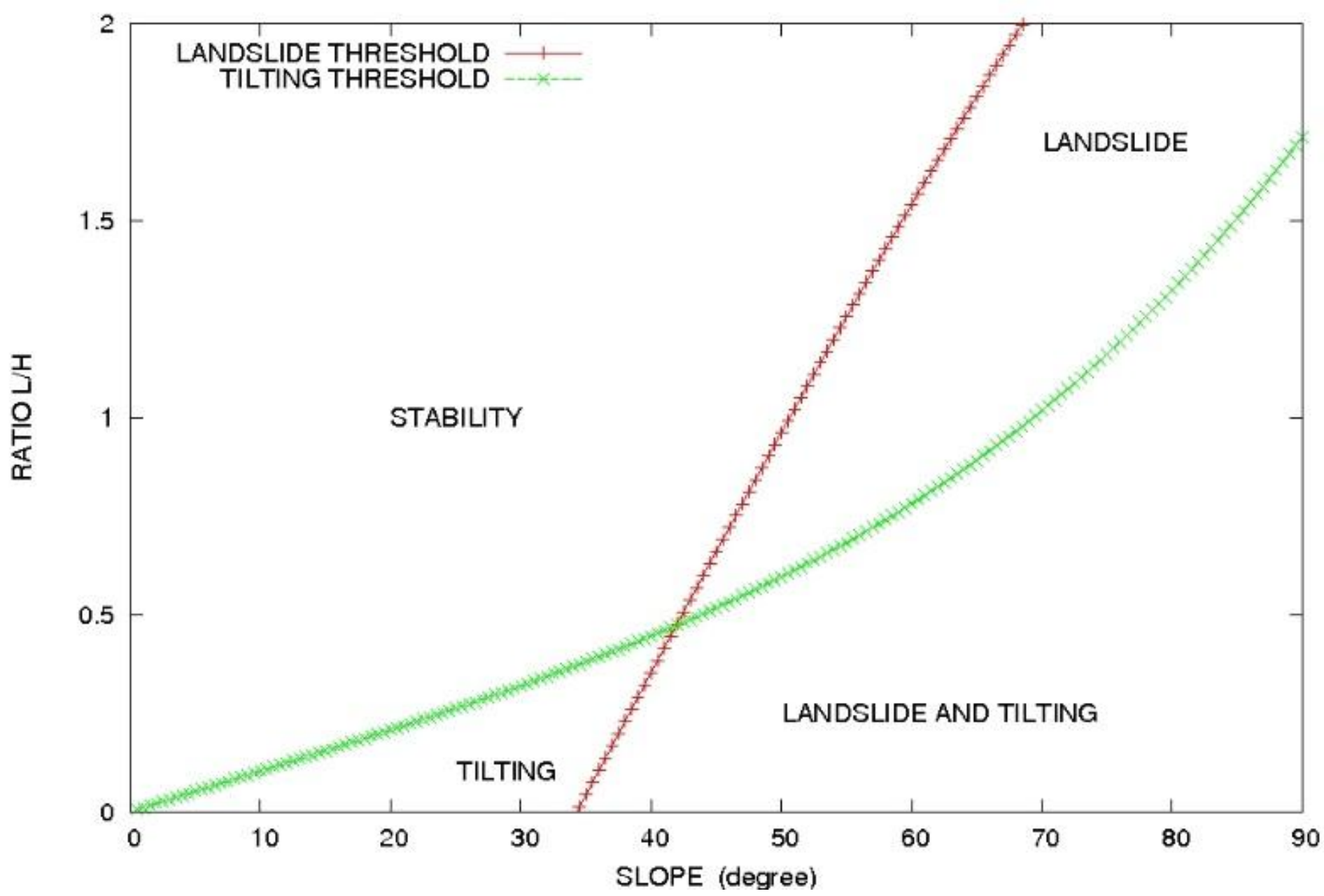

**Figure 13.** Distinction between the main destabilization process that take place using the geomechanical thresholds. The ratio between the length L and the height of escarpment $H$ as a function of the slope $\theta$. $C$=10MPa, $T$=7.8 MPa, $\phi$=34.3°, $L$=1000m, $g$=9.8ms$^{-2}$, $\rho$=3000 kg/m$^3$.

From a practical point of view, it is often possible to distinguish between processes based on field study [40,73] or by remote sensing [74,75] for relatively recent events using accurate topographical data. However, the task becomes considerably more difficult for older morphologies or when data resolution is limited [34]. In some cases, a combination of topographic and geometrical analyses (e.g., the presence or absence of faults, the presence of water outlets, or the spatial distribution of debris) can provide critical insights on the most plausible destabilization scenario [76,77]. In other cases, the distance between the initial and final positions of the debris can provide clues about the dynamic of the event, allowing differentiation between landslides, rock falls, or tipping mechanisms [4,73,78,79].

However, significant uncertainties may exist on the thresholds, not only caused by the kind of destabilization process, but also in relation to the mechanical law considered. This is the case when tilting is the main destabilization process, but this is also the case when sliding is the main destabilization process. The maximum height threshold of the escarpment, when landslide is the main destabilizing process, differs considerably depending on whether one used a model based on the Mohr-Coulomb law, or a model based on the Griffith criterion with a constant traction $T$, or a model based on the Griffith criterion with a variable traction $T$ (Figure 14). Applying mechanical laws implies certain consequences on the maximum relief for a given slope. Basically, in the case where landslides constitute the main destabilization process, the maximum height of the escarpment are:

$H_{MC}^{S} = C / [\rho g (\sin\theta - \cos\theta \tan\phi)]$, assuming the Mohr-Coulomb law,

and $H_{G}^{S} = [2T \cos\theta + 2T] / [\rho g \sin^2\theta]$ assuming the Griffith criterion.

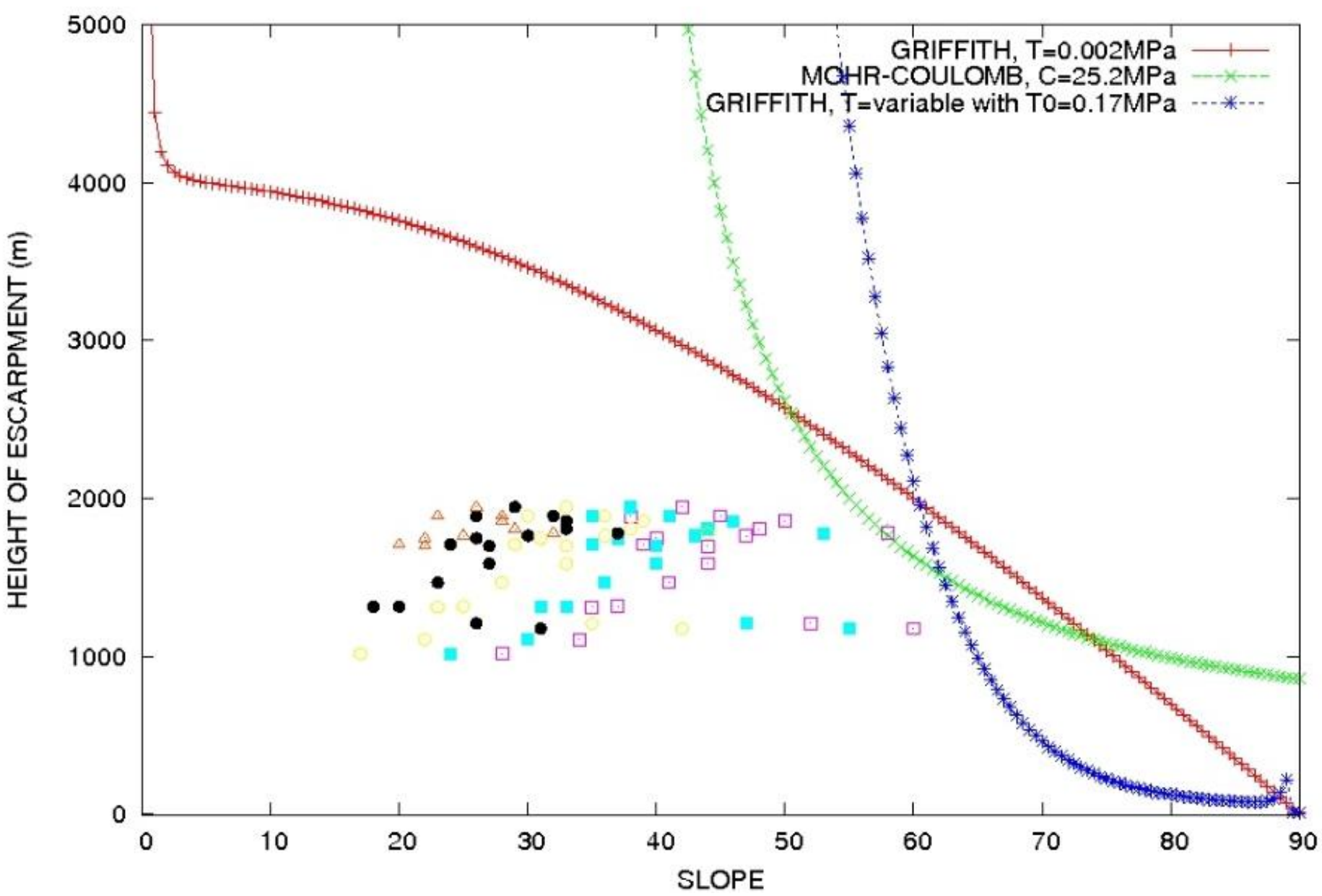

**Figure 14.** Landslide thresholds. Maximum height of escarpments as a function of slope in the case where landslide is the unique slope destabilization process using : (i) the Griffith criterion with a constant traction $T$=0.002MPa to estimate $H_G{}^S$ represented by a red curve, (ii) the Griffith criterion with a variable traction as in equation (13) to estimate $H_G{}^S$ represented with a blue curve, and (iii) the Mohr-Coulomb law with $C$ = 25.2 MPa, $\phi$ = 34.3° to estimate $H_{MC}{}^S$ represented by a green curve. In all cases the other parameters are $\rho$ = 3000 kg/m³, $g$=9.8ms⁻², $L$ = 1000m. Concerning the variable traction $T$ the parameters values are $a$=5, $b$=3 MPa, $c$=1.5.

Another limit of this theoretical approaches lies in the conditions that lead to the formation of the relief. The relief elevation could be limited by uplift force amplitude. The theoretical maximum height of an escarpment could be higher than the observed relief in certain cases, if the amount of tectonic force was not locally sufficient to reach the maximum potential height of the escarpment. For example, in the Carrara area, the Apennines relief is less than 2 000m, and this altitude may have been limited by a restricted vertical movement, during the last million years. Vertical movements and reliefs in the Apennines were caused by the collision-subduction boundary and the slab rollback [80-84]. The present uplift rate in the Apennines ranges from 0.2 to 1 mm/yr [85,86].

*4.2. Discriminating between the best failure criterion at the mountain range scale*

The validation of theoretical models generally relies on (i) model comparison –an approach used for example for global climate models (GCM) [87,88]–, (ii) extrapolation from laws validated at different scales and under different conditions [89-91], (iii) the capacity to predict real events [92], (iv) the selection of scenarios considered more relevant according to the geological context [93-97].

The Mohr-Coulomb law is a classical approach to geotechnical analysis used to estimate the safety factor $F_S$ for landslides or rock block toppling [43,44,45,98,99]. The Mohr-Coulomb law provides estimates of the maximum height of escarpments when landslides or block tilting are the main slope destabilizing processes. The results obtained in this study using the More-Coulomb law are in agreement with the results obtained in previous studies using the same approach [34].

However, the new equation used in this study to model the failure envelope is different from the Mohr-Coulomb law and from the Hoek-Brown failure envelope [100], but describes better the tensile domain than the Mohr-Coulomb equation. The approach developed here is also different from the conventional approach resulting from the Griffith criterion, because the traction $T$ is considered variable in the present case. Despite the equation used in this study to describe the mechanical behavior is different from the one used by Li and Ghassemi (2019) [51], its general shape is almost similar.

The Griffith law with variable traction yields four theoretical solutions for the maximum height of escarpments caused by tilting (called $H_G$). One of these solution is negative and thus physically irrelevant for escarpment height. Eventually, this solution could be reinterpreted in the case of a depression, if the vertical reference frame is adjusted. Among the remaining solutions, two are equal, leaving only two distinct values. However, the physical meaning of one of these solutions (Figure 10, solution 3, green curve) is questionable: it is independent of traction $T$ and predicts that rocks should tilt when the slope reaches 90°, regardless of rock resistance. The empirical observation that not all vertical slopes collapse or tilt suggests that this solution is, at best, not universally applicable. Finally, only one solution seems to be relevant to quantify slope stability.

Although the Mohr-Coulomb approach has been widely used for decades to assess rockfall risks, our results suggest that the Mohr-Coulomb law may systematically overestimate the maximum height of escarpments. This overestimation is often mitigated in practice by the introduction of safety coefficients that adjust effectives cohesion and friction angle. This practice is justified by the need to account for rock weathering and the heterogeneity of rocks. From a theoretical perspective, however, this difference could be also caused by the potential role of hybrid and tensile fractures in weakening of rocks. Neglecting these types of fractures, may lead to overestimate rock resistance. Tensile and hybrid fractures have been observed in various lithology such as limestone, sandstone [51] and marble [57]. Therefore, this type of approach can be applied for various lithologies and contexts. From a purely theoretical point of view, the results suggest that hybrid and tensile fractures, taken into account in this study, play a role in the maximum height of escarpments and, consequently, on the overall morphology of the relief.

The difficulty to assess the relevance of the Griffith criterion with a variable traction $T$ to describe the high relief, could be caused by the potential perturbation from seismic and climatic conditions on slope stability, as well as by the superposition of many destabilization processes (landslide, tilting, water erosion, glacial erosion, debris flow,wind erosion). These perturbations may be of the same order of magnitude as the uncertainties arising from the choice of the mechanical law (e.g., Mohr-Coulomb vs. Griffith).

## 5. Conclusions

The main conclusions of this study are as follows:

1. A new failure criterion is necessary to accurately describe tensile, hybrid and shear fractures, and an equation is proposed in this study;
2. The rock traction parameter can be variable in hybrid and tensile domain;
3. The use of this new failure criterion results in a different stability threshold for the maximum height of escarpments caused by tilting compared to the use of the Mohr-Coulomb law;
4. The Mohr-Coulomb law tends to overestimate the maximum height of escarpment compared to the Griffith criterion;
5. Considering tensile and hybrid fractures has an effect on the maximum height of escarpments of the same order of magnitude as the potential influence of a triggering event, such as the seismicity or significant pore pressure;
6. The effect of considering tensile and hybrid fractures on the maximum height threshold is of the same order of magnitude as the main destabilization process considered (sliding vs. tilting).

The findings of this study have significant implications for natural hazard assessment and geotechnical practice. The systematic overestimation of escarpment heights by the Mohr-Coulomb law, if uncorrected, could lead to underestimation of risk in regions where tensile or hybrid fractures dominate. Conversely, the Griffith's criterion, while more accurate, introduces additional complexity. Future research should focus on integrating high-resolution topographical data, mechanical testing of heterogeneous rocks, and probabilistic approaches to better constrain the uncertainties associated with physical laws and environmental forcing. In doing so, more robust predictions of slope instabilities and their cascading hazards, will improve risk mitigation strategies in vulnerable mountain regions.